\documentclass{rnaastex}
\usepackage{amsmath,color,textcomp,url,graphicx,subfigure}
\usepackage{morefloats}

\newcommand{\kms}{\hbox{km\,s$^{-1}$}}
\newcommand{\mjup}{$M_{\mathrm{Jup}}$}
\newcommand{\masyr}{$\mathrm{mas}\,\mathrm{yr}^{-1}$}

\begin{document}

\title{A GAIA DR2 CONFIRMATION THAT 2MASS~J12074836--3900043 IS A MEMBER OF THE TW~HYA ASSOCIATION}

\author[0000-0002-2592-9612]{Jonathan Gagn\'e}
\affiliation{Carnegie Institution of Washington DTM, 5241 Broad Branch Road NW, Washington, DC~20015, USA}
\affiliation{NASA Sagan Fellow}
\email{jgagne@carnegiescience.edu}
\author[0000-0003-4636-6676]{Eileen C. Gonzales}
\affiliation{Department of Astrophysics, American Museum of Natural History, Central Park West at 79th St., New York, NY 10024, USA}
\affiliation{The Graduate Center, City University of New York, New York, NY 10016, USA}
\affiliation{Department of Physics and Astronomy, Hunter College, City University of New York, New York, NY 10065, USA}
\altaffiliation{LSSTC Data Science Fellow}
\author[0000-0001-6251-0573]{Jacqueline K. Faherty}
\affiliation{Department of Astrophysics, American Museum of Natural History, Central Park West at 79th St., New York, NY 10024, USA}

\keywords{methods: data analysis --- stars: kinematics and dynamics --- proper motions}

\section{}

The low-gravity L1\,$\gamma$ substellar object 2MASS~J12074836--3900043 (2M1207--3900 hereafter) was discovered by \cite{2014ApJ...785L..14G} as an extremely low-mass candidate member of the TW~Hya association (TWA). The model-dependent mass estimate of 11--13\,\mjup\ is near the deuterium-burning boundary. Two lower-mass high-likelihood members were more recently discovered \citep{2016ApJ...822L...1S,2016ApJ...821L..15K}, but they still lack a trigonometric parallax measurement for full membership confirmation.

The membership assessment of 2M1207--3900 was based on only proper motion and sky position at the time, which were analyzed with the BANYAN~II tool \citep{2014ApJ...783..121G} to yield a Bayesian membership probability of 99.6\%. \cite{2017ApJS..228...18G} later published a radial velocity of $6 \pm 3$\,\kms, further strenghtening its membership, maintaining a Bayesian probability of 99.6\%, with a predicted kinematic distance of $59.8 \pm 5.6$\,pc if it is a member of TWA.

The recent release of \emph{Gaia}~DR2 \citep{Lindegren:2018gy} reported a proper motion of $\mu_\alpha\cos\delta = -61.5 \pm 1.4$\,\masyr, $\mu_\delta = -23.9 \pm 0.8$\,\masyr\ for 2M1207--3900, and a trigonometric parallax measurement of $14.9 \pm 0.9$\,mas, which approximately corresponds to a distance of $67.1 \pm 4.0$\,pc, $
\sim$\,1$\sigma$ further away than the BANYAN~II prediction. Combining these measurements with the \cite{2017ApJS..228...18G} radial velocity in the updated tool BANYAN~$\Sigma$ \citep{2018ApJ...856...23G} yields a TWA membership probability of 99.8\%, and places 2M1207--3900 at a kinematic separation of 2.9\,\kms\ from the core of the TWA members in $UVW$ space. We therefore suggest that 2M1207--3900 is a new bona fide member of the TWA association.

We used the updated trigonometric distance from \emph{Gaia}~DR2 with the method of \citeauthor{2016ApJS..225...10F} (\citeyear{2016ApJS..225...10F}; see also \citealt{2015ApJ...810..158F}) to update the model-dependent mass estimate of 2M1207--3900: we obtain a mass of $15.3 \pm 1.2$\,\mjup, at the very low-mass end of the substellar regime. The revised larger mass is a consequence of its trigonometric distance which is slightly larger than the kinematic distance predicted by BANYAN~II.

\cite{2014ApJ...785L..14G} also found the young M9\,$\gamma$ brown dwarf 2MASS~J12474428--3816464 (2M1247--3816 hereafter) as a lower-probability candidate member of TWA (29.4\%), but its membership was highly uncertain given that its kinematics were slightly discrepant from that of other TWA members. \emph{Gaia}~DR2 provided updated kinematics for this object as well ($\mu_\alpha\cos\delta = -43.4 \pm 0.9$\,\masyr, $\mu_\delta = -22.1 \pm 0.6$\,\masyr, with a trigonometric parallax of $11.8 \pm 0.5$\,mas). New data for 2M1247--3816 from the Folded-port InfraRed Echellette spectrograph (FIRE; \citealp{2008SPIE.7014E..0US,2013PASP..125..270S}) at the Magellan Baade telescope yielded a radial velocity measurement of $1.7 \pm 3.8$\,\kms\ for this object (J.~Gagn\'e et al., in preparation; see the radial velocity measurement method described in \citealt{2017ApJS..228...18G}). Including these updated kinematics in BANYAN~$\Sigma$ yields a higher 87.9\% Bayesian membership probability in TWA, but places 2M1247--3816 at 4.6\,\kms\ from the kinematic core of other TWA members. It is worth noting that this object obtains a 0\% Bayesian membership probability in the Lower Centarus Crux association, which is more distant than TWA and has been known to contaminate searches for new TWA members. It remains unclear at this time whether 2M1247--3816 is a genuine member of TWA, or if it is a young interloper. It would also be one of the most distant members of TWA at $85 \pm 3$\,pc. If this object is unrelated to TWA, it is unclear where it would have formed, as is the case for several other young brown dwarfs with full kinematics (e.g., see \citealt{2016ApJS..225...10F}).

\acknowledgments

This research made use of: data products from the Two Micron All Sky Survey, which is a joint project of the University of Massachusetts and the Infrared Processing and Analysis Center/California Institute of Technology, funded by NASA and the National Science Foundation, and data from the ESA mission \emph{Gaia}, processed by the {\it Gaia} Data Processing and Analysis Consortium whose funding has been provided by national institutions, in particular the institutions participating in the {\it Gaia} Multilateral Agreement. This research note was written at the 2018 \emph{Gaia}~DR2 workshop, hosted by the Flat Iron Institute in New-York city.

\facility{Magellan:Baade (FIRE)}
\software{BANYAN~II \citep{2014ApJ...783..121G}, BANYAN~$\Sigma$ \citep{2018ApJ...856...23G}.}


\end{document}